\documentclass{svproc}
\usepackage{url}

\usepackage{listings}
\usepackage{newfloat} 

\DeclareFloatingEnvironment[
    fileext=lol,
    listname=List of Listings,
    name=Listing,
    placement=tbhp
]{listing}

\usepackage{tabularx}
\usepackage{makecell}

\usepackage{cite}
\usepackage{multirow}
\usepackage{amsmath,amssymb,amsfonts}
\usepackage{algorithmic}
\usepackage{graphicx}
\usepackage{textcomp}
\usepackage{xcolor}
\usepackage{url}
\definecolor{siamak}{rgb}{1, 0, 0}
\definecolor{yaying}{rgb}{0.58,0,0.82}
\definecolor{codegreen}{rgb}{0,0.6,0}
\definecolor{codegray}{rgb}{0.5,0.5,0.5}
\definecolor{codepurple}{rgb}{0.58,0,0.82}
\definecolor{backcolour}{rgb}{0.98,0.98,0.98}

\lstdefinestyle{mystyle}{
    backgroundcolor=\color[RGB]{247, 247, 247},   
    commentstyle=\color{codegreen},
    keywordstyle=\color{magenta},
    numberstyle=\tiny\color{codegray},
    stringstyle=\color{codepurple},
     basicstyle=\fontfamily{pcr}\fontsize{8.3}{14}\selectfont\linespread{1.3},
    breakatwhitespace=false,         
    breaklines=true,                 
    captionpos=t,                    
    keepspaces=true, 
    xleftmargin=2em,
    framexleftmargin=2em,
    numbers=left,                    
    numbersep=3em,                  
    showspaces=false,                
    showstringspaces=false,
    showtabs=false,
    tabsize=2,
        frame=tb,           
    framerule=0.5pt     
}
\def\BibTeX{{\rm B\kern-.05em{\sc i\kern-.025em b}\kern-.08em
    T\kern-.1667em\lower.7ex\hbox{E}\kern-.125emX}}
\begin{document}
\sloppy

\title{P4-NIDS: High-Performance Network Monitoring and Intrusion Detection in P4}



\author{Yaying Chen\inst{1} \and Siamak Layeghy\inst{1} \and Liam Daly Manocchio\inst{1} \and Marius Portmann\inst{1}}

\titlerunning{P4-NIDS, Intrusion Detection in P4}

\authorrunning{Yaying Chen et al.}

\institute{School of EECS, The University of Queensland, Brisbane, QLD, 4072, Australia \\
\email{yaying.chen@uq.net.au, siamak.layeghy@uq.net.au, liam@riftcs.com and marius@ieee.org}}


\maketitle

\begin{abstract}
This paper presents a high-performance, scalable network monitoring and intrusion detection system (IDS) implemented in P4. The proposed solution is designed for high-performance environments such as cloud data centers, where ultra-low latency, high bandwidth, and resilient infrastructure are essential.
Existing state-of-the-art (SoA) solutions, which rely on traditional out-of-band monitoring and intrusion detection techniques, often struggle to achieve the necessary latency and scalability in large-scale, high-speed networks. Unlike these approaches, our in-band solution provides a more efficient, scalable alternative that meets the performance needs of Terabit networks.
Our monitoring component captures extended NetFlow v9 features at wire speed, while the in-band IDS achieves high-accuracy detection without compromising on performance. In evaluations on real-world P4 hardware, both the NetFlow monitoring and IDS components maintain negligible impact on throughput, even at traffic rates up to 8 million packets per second (mpps). This performance surpasses SoA in terms of accuracy and throughput efficiency, ensuring that our solution meets the requirements of large-scale, high-performance environments.
\end{abstract}


\section{Introduction}

High-performance networks should be able to transfer vast amounts of data with very minimal delays, usually in the order of microseconds, so that real-time applications in areas such as  cloud computing are realised. 
They must meet strict requirements to support modern applications and services, including ultra-low latency, high bandwidth, scalability, and resilience~\cite{kiran2018enabling}.
In addition, these networks need to maintain consistent performance under varying load conditions and scale effectively  to handle increasing traffic demands. Robust security and intrusion detection are essential to protect sensitive data and ensure operational integrity against ongoing cyber threats. Equally important is efficient resource management, achieved through advanced monitoring to optimise  performance and resource allocation~\cite{kiran2019understanding}.

To meet these requirements, advanced monitoring techniques are essential for keeping high-performance networks running smoothly. Real-time visibility allows for the immediate identification and response to potential issues before they impact users. This includes monitoring network traffic patterns, device performance, and application behaviour~\cite{marques2019optimization}.
Telemetry data from various network elements provides valuable insights for capacity planning and performance optimisation. Additionally, end-to-end monitoring across the entire network path aids in troubleshooting complex issues in distributed environments~\cite{oberdorf2023predictive}.

Similarly, intrusion detection plays a critical role in enforcing security measures. Modern intrusion detection techniques go beyond simple port-based methods, which only check port numbers assigned to specific applications. Instead they incorporate Deep Packet Inspection (DPI) and machine learning (ML) algorithms to accurately identify application types and traffic patterns. 
Traffic flows, in particular, are widely used for network traffic analysis to extract information related to applications based on their network behaviour. However, in high-performance networks, real-time intrusion detection that enables dynamic policy enforcement and mitigates security threats remains a challenging problem~\cite{salman2020review, setiawan2022encrypted}.


P4 (Programming Protocol-independent Packet Processors)~\cite{p4} is a domain-specific language designed to specify how packets are processed by the data planes of network forwarding elements. Its introduction  brought a revolutionary approach to networking and network management by enabling the programming of data planes~\cite{hauser2023survey}. This programmability empowers network operators to define custom packet processing logic directly within network devices, unlocking new possibilities for fine-grained traffic analysis and adaptive monitoring. 
An additional advantage of P4 is its protocol independence, ensuring that solutions developed with it are future-proof, capable of adapting to new network protocols and evolving traffic patterns. This makes P4 a strong foundation for achieving high performance in network management.

While P4 offers significant advantages for monitoring and intrusion detection, it also has several limitations~\cite{das2023memory}. First, the memory and processing capabilities available in hardware switches for P4 are limited~\cite{gonccalves2019random}, which restricts the complexity of monitoring and classification algorithms that can be implemented directly in the data plane.
As a result, more complex analysis often needs to be offloaded to external systems, which can introduce additional latency and reduce real-time responsiveness. Secondly, since P4 focuses on packet-level processing, maintaining state across multiple packets-essential for many advanced intrusion detection techniques-becomes challenging~\cite{pekar2024p4tonfv}.


This research proposes a P4-based monitoring and intrusion detection solution for high-performance networks, designed to operate within SDN data plane limitations.
Our monitoring solution captures NetFlow~\cite{netflow} data with enhanced v.9 features, not found in previous solutions, and runs at wire speed, making it ideal for Terabit networks.
We also implement a lightweight, flow-based intrusion detection system, as proposed in~\cite{lightbulb}, within P4. This system uses NetFlow data from our monitoring module, addressing the need for cross-packet state tracking.

The main contributions of this research can be summarised as:

\begin{itemize}
\item 
    \textbf{Development of a high-performance, in-band network monitoring solution in P4} 
    
    We design and implement a NetFlow exporter capable of generating advanced NetFlow v.9 fields, which is not typically available in standard solutions. This exporter operates at wire speed and is capable of providing real-time network traffic visibility, offering a more scalable and efficient alternative to traditional out-of-band solutions.
\item 
    \textbf{In-band implementation of a lightweight and efficient NIDS in P4}
    
    Our NIDS solution operates directly within the data plane, independent of the SDN controller, offering low-latency, high-throughput security monitoring. This approach improves real-time detection capabilities without compromising overall network performance.
\item 
    \textbf{Comprehensive performance evaluation in a 40 Gbps P4 hardware}
    
    We assess both the NetFlow exporter and the NIDS in terms of computational overhead and impact on throughput. Our solution is capable of processing traffic rates up to 8 million packets per second (mpps) with minimal impact on throughput, even when both modules are active. By comparing our approach with state-of-the-art solutions, we demonstrate the superior efficiency and scalability of our solution for high-speed networks.
\end{itemize}

The rest of this paper is organised as follows: Section~\ref{Background} provides an overview of P4 programming and the NetFlow format. Section~\ref{Related Work} discusses previous research in the fields of in-band network monitoring and intrusion detection, analysing their strengths and limitations. The following section outlines the methodology for in-band NetFlow generation and NIDS implementation in P4. Section~\ref{Evaluation and Results} presents an evaluation of the proposed solutions, first in an emulated network environment and then on a 40 Gbps P4 hardware setup, with results for each experimental configuration. Finally, Section~\ref{Conclusion} summaries the key findings and concludes the paper.

\section{Background} \label{Background}

\subsection{P4 Programming}
The Programming Protocol-independent Packet Processors (P4) is an open-source SDN data plane programming language that was introduced in 2014 with the aim of providing reconfigurability in the field, protocol independence, and target independence.
This indicates that P4 programmers should have the capability to modify how switches handle packets after they are deployed. It suggests that switches should not be restricted to particular network protocols, and programmers should have the freedom to define packet-processing functions regardless of the complex details of the hardware they are built upon~\cite{p4}. 

Through the separation of packet processing logic from hardware, P4 empowers devices to accommodate diverse networking protocols and provides developers with the flexibility to adjust device behavior without hardware modifications. This decoupling promotes innovation and adaptability in networking, enabling developers to explore new protocols, enhance network performance, and create tailored traffic management policies. Additionally, P4 facilitates the prototyping and implementation of novel networking features, enabling rapid iteration on designs without extensive hardware alterations. This speeds up development timelines and facilitates quicker deployment of advanced networking functionalities and services, providing more agile and effective network infrastructures.

 The \textit{P4 ecosystem} provides a comprehensive suite of tools, compilers, libraries, and targets for the development, testing, and deployment of P4-based network functions. There are some critical components within that ecosystem and the language itself, which play a very important role in materializing programmable network infrastructure. These key elements are explained as follows:

\begin{itemize}
 \item{\textbf{P4 Language:} At the core is the P4 language, which provides a protocol-independent and very flexible way of specifying packet processing behavior to network engineers or developers. P4 abstracts the hardware specification and hence makes it possible to program network equipment for specific use cases and requirements.}

\item{\textbf{P4 Compiler:} The P4 compiler, developed at ETH~\cite{ETH}, is a member of the family of portable compilers that translate high-level P4 programs into lower-level, target-specific implementations. Optimizations are created, and configurations are generated for specific devices in order to guarantee efficient packet processing. P4 conformance, which means conformance of a target to the language specification with respect to syntax, semantics, and features of P4, is quite important for the correct interpretation and execution of P4 programs.}

\item{\textbf{P4 Runtime Environment:} P4Runtime is a protocol-agnostic API that allows field reconfigurability. In particular, it offers the following abilities: dynamic deployment of a new P4 forwarding plane onto devices without needing code recompilation~\cite{p4Runtime}. Network operators can update packet processing logic dynamically without any disruption to operations.}

\item{\textbf{P4 Targets:} These are the platforms (hardware or software) on which P4 programs can be run. They include programmable Application-Specific Integrated Circuits, FPGA-based SmartNICs, software switches like Open vSwitch, and Programmable Network Processors~\cite{p4target}.}

\item{\textbf{P4 Tools and Utilities:} These are a raft of resources to help develop and debug P4 programs provided by P4 community. These range from debuggers, packet trace analysis tools, libraries,  frameworks that enable many common networking tasks and protocols using the P4 language and  simulation software.}
\end{itemize}

\subsection{NetFlow}
Network management and security require the collection and analysis of network traffic. There are three main approaches to network data collection.
At one end of the spectrum is the collection of high-level information related to network device management. An example of this approach is the Simple Network Management Protocol (SNMP), an internet standard for monitoring and managing network devices.
SNMP allows network administrators to gather information about network devices, such as routers, switches, servers, and printers, by polling these devices for various metrics such as bandwidth usage, interface errors, and CPU load. This approach provides a basic overview of network health and performance but lacks the detailed information needed for many network security and troubleshooting operations.

On the other end of the spectrum is packet capturing, where all traffic packets passing through monitoring points are collected and recorded. While this method provides much more detailed and useful information compared to the first approach, it is almost impractical in large-scale real-world networks due to the storage requirements and concerns regarding the privacy and security of network users and servers.

The third approach involves summarising network traffic statistics as network flows, balancing between the detailed but storage-intensive packet capturing and the high-level statistics of SNMP-like protocols. Network flows are identified by common attributes known as the five-tuple, including source and destination IP addresses, L4 (transport layer) ports, and the L4 protocol, representing a series of packets. This approach is widely adopted in the networking industry for its scalability and practicality.

Accordingly, all major producers of network devices have developed or adopted methods for network flow collection. For example, Huawei uses the NetStream protocol, Juniper uses Jflow, Alcatel uses cflow, and s-flow is employed by several large network device manufacturers. However, the most widely adopted network flow collection protocol is Cisco's NetFlow, proposed in 1996 by Darren and Barry Bruins~\cite{cisco}. 
NetFlow is also the basis for an Internet Engineering Task Force (IETF) standard for flow information, known as Internet Protocol Flow Information Export (IPFIX). There are different versions of the NetFlow protocol that provide varying levels of information about network traffic. While NetFlow version 5 is unidirectional with a limited set of features, NetFlow version 9, the most common version, offers bi-directional flow information with a much broader set of features~\cite{Cisco2011}.

\section{Related Work} \label{Related Work}
\vspace{-0.35cm}
Since the proposed method for intrusion detection in this research is flow-based, and we have also developed a method for in-band NetFlow generation/extraction, we review the related works in both fields. First, we focus on prior approaches to in-band NetFlow generation, then we examine existing in-band flow-based intrusion detection systems, highlighting the techniques and models they employ to establish the context and limitations that our work addresses.

\subsection{In-band Network Monitoring}\label{Network Monitoring}

FlowStalker~\cite{castanheira2019flowstalker} is proposed to enhance the ability of monitoring traffic flows directly on the data plane,  without relying on external controllers or middleboxes.
The main idea is to leverage P4 to monitor network flows directly on the data plane. FlowStalker allows for capturing a wide range of metrics, including packet counts, byte counts, and a few other flow statistics.
%
%
While FlowStalker brings flexibility to directly monitor a wide range of traffic metrics on the data plane, it does so at an extra cost with respect to management and optimisation of the monitoring rules. This extra complexity is more critical in large-scale network environments that are dynamic in nature, with traffic patterns constantly changing with a high number of flows. 
Additionally, FlowStalker has not been evaluated on real-world hardware, as the authors relied exclusively on BMv2~\cite{BMV2}, a P4 software switch that does not fully capture the constraints and performance characteristics of production-grade devices. This gap leaves open questions about how well FlowStalker would scale under the resource limitations and high-throughput demands of actual hardware environments.

P4Flow~\cite{mostafaei2021p4flow}, proposes a flow monitoring tool for cloud provider networks implemented on programmable data planes. It allows the providers to monitor a set of desired flows according to their needs, offering more granular control over what is monitored and how data is collected. The system is designed to operate without the need for extensive controller intervention, enabling monitoring directly on the network switches.
While it is claimed that P4Flow can reduce the overhead of reporting the flow statistics by considering the properties of the inter-node links such as delay and bandwidth, they have not included the effect of monitoring flows via P4Flow on throughput or other resources.
%
%
Although P4Flow is powerful in monitoring network traffic, it comes at a high resource cost in terms of memory, processing power, and other resources on the switch, which might be exhausted and hence limit its deployment, especially on resource-constrained environments. Network operators need to manage these resources cautiously, probably at the cost of monitoring details versus total network performance, to let P4Flow be effectively deployed without overwhelming the underlying hardware.

P4-based Proactive Monitoring (PPM)~\cite{oh2024p4} adopts a proactive approach to enhance the efficiency of monitoring collection operations. 
It collects and processes network data in real time directly within the network devices. This is in contrast to traditional reactive monitoring schemes that rely on the central controller to request data from switches.
PPM allows programmable switches to proactively forward monitoring information to the controller after the controller enables PPM. The measurement results show that PPM can not only enhance the efficiency of collecting monitoring information by applying a proactive mechanism but also minimise the general monitoring overhead compared to the polling-based methods.
However, PPM still requires the SDN controller as coordinator and while the scheme reduces direct dependence on the controller, the controller still plays an essential role as a coordinator. It can configure the P4 programs deployed on the switches and set policies that dictate how monitoring should be conducted. In addition, the performance evaluation with realistic traffic rates is missing in this research.

\subsection{In-band Intrusion Detection}
IISY~\cite{xiong2019switches} explores using commodity programmable switches for in-network classification. While it is not designed as an NIDS, it could feasibly be adapted for this purpose. 
It focuses on mapping trained machine learning models to match-action pipelines, and introduces a prototype named In-network Inference made easy(IIsy). 
Support for the integration of four machine learning models at the data plane level: decision trees, support vector machines, K-means, is illustrated. 
The authors also discuss the adaptability of their approach to various targets and suggest that their solution can be extended to include additional machine learning algorithms.
However, the approach is inherently limited in accuracy due to the types of features it can extract. This restriction can impact the reliability of the classification process.
Also, deploying updates to classification models requires changes through the control plane, which could add complexity and delay in adapting to new requirements or improving model performance.

Researchers at Purdue University proposed Leo~\cite{leonsdi2024} for online intrusion detection capable of handling multi-terabit line rates.
It employs a machine learning model that operates within the data plane of network switches. 
In order to achieve higher speeds, Leo restricts the number of features used for packet classification to 10, and this limitation affects the accuracy of classification for complex traffic patterns.
In addition, while Leo supports deeper trees than many previous systems (up to depth in TCAMs), it still relies on balancing between tree depth and accuracy, which could lead to performance issues if the depth is insufficient for certain classification tasks.
In summary, while Leo presents a promising approach to online intrusion detection, its limited classification accuracy highlights the need for alternative solutions in this field.


MAP4~\cite{xavier2022map4} is a framework designed to explore the feasibility of deploying ML models in programmable network devices. It deploys a pre-trained Decision-Tree model into a programmable switch, allowing for accurate flow classification at line rate. The results demonstrate that most flows can be correctly classified with just a few packets. In some scenarios, 97\% of the traffic is accurately classified with only two packets per flow, while all traffic classes are correctly labelled using a maximum of four packets.
However, the MAP4 framework requires a concurrency control mechanism implemented in Micro C language to prevent race conditions arising from parallel packet processing. This reliance on an external component introduces a dependency outside the P4 ecosystem and could limit portability across different hardware platforms. 

In order to address the shortcomings of the previous systems, as explained above, this paper proposes an approach for both network monitoring and intrusion detection, implemented purely in P4, with no use of the SDN controller or reliance on any other languages.

\section{Method} \label{Method}
Figure~\ref{fig:P4Switch} shows the overall architecture of the proposed solution for both hardware-based and emulation-based implementation. 
The left figure shows the implementation in a P4-compatible hardware data plane, and the right figure shows the implementation in BMv2 emulation environment. 
As can be seen, in both implementations there are two main  components: a \textit{NetFlow generator/exporter} and a pre-trained ML model. 
In this section, we initially explain the NetFlow generator/exporter implementation, then we discuss the implementation of the pre-trained ML model.

{\begin{figure}[!t]
	\centering
	\includegraphics[width=1\linewidth]{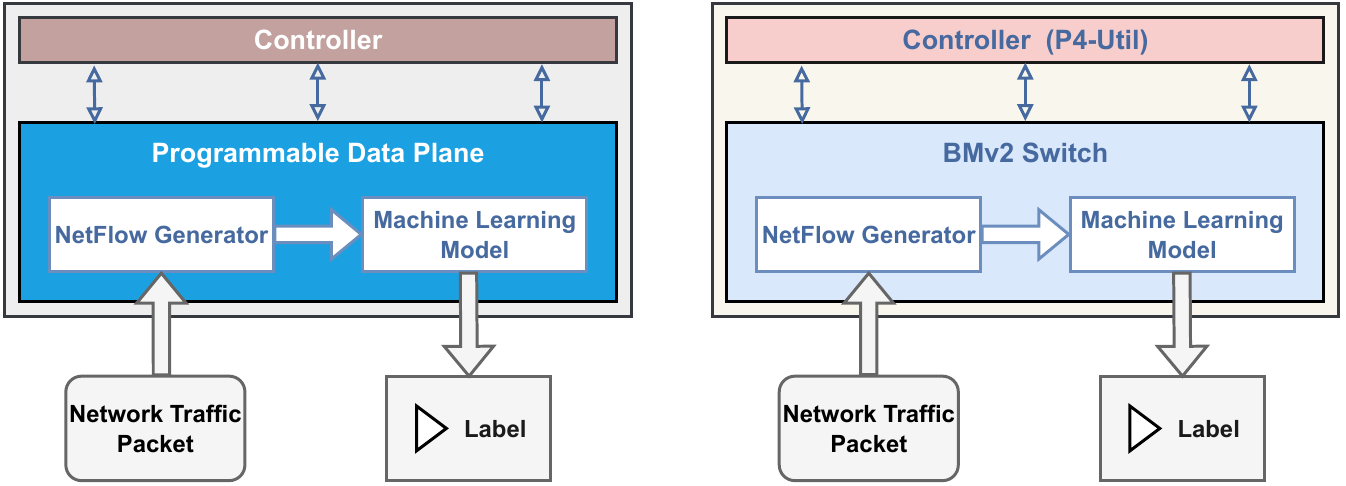} 
	\caption{P4-NIDS architecture in left) Hardware data plane and Right) BMv2 emulation environment}
        \label{fig:P4Switch}
\end{figure}}

\subsection{NetFlow Generator in P4}


{\begin{figure}[!b]
	\centering
	\includegraphics[width=1\columnwidth]{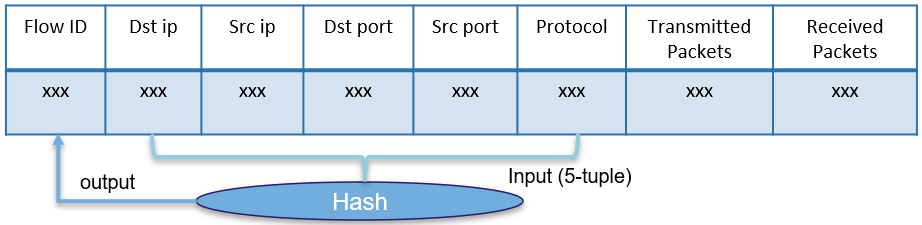} 
	\caption{NetFlow Table, and fields used to create the hash value for a flow entry}
        \label{fig:NetFlowTable}
\end{figure}}

Upon arrival at a switch port, each packet is processed by the NetFlow Generator component, which logs flows using a Flow ID. This Flow ID is derived from applying a hash function to the packet’s 5-tuple (destination IP, source IP, destination port, source port, and protocol). The Flow ID serves as a key for accessing entries in the NetFlow Table (Figure~\ref{fig:NetFlowTable}) within the P4 Switch. This table is organised as an array of registers, where each unique NetFlow field is stored in its own designated P4 Register. 
When flows are analysed for intrusion detection, e.g., by an ML-based NIDS, these registers can be accessed by the ML model directly on the data plane. Currently, the system supports 12 NetFlow fields~\cite{SARHAN2022100359} , enabling the necessary features for an ML-based NIDS. Additional NetFlow fields could be generated similarly if required.

Furthermore, these registers are accessible from the control plane using the P4-Utils~\cite{p4utils} API, allowing periodic exports as CSV files or transmission to an external NetFlow collector server. This module can therefore function independently as a P4-implemented NetFlow Exporter.

\subsection{ML-based NIDS in P4}
The P4 language is designed mainly for packet processing but has several limitations that make it difficult to implement most machine learning models. P4 lacks general-purpose computation features like variables, loops, and functions. It also has limited memory and stateful processing because it relies on basic register-based storage, and it does not support floating-point arithmetic or complex data types, which are essential for many machine learning tasks. Furthermore, P4’s control flow is not flexible enough to handle the complex logic needed for machine learning models, such as recursive functions or deep learning architectures. The absence of built-in machine learning libraries adds to the challenge, requiring developers to implement fundamental algorithms from scratch.~\cite{hauser2023survey}.

While P4 has limitations, it is still possible to implement certain models, such as Decision Tree classifiers, Random Forests, and simple regression models using its available instruction set. While simple neural network models have also been implemented in P4 environments, this was made possible by extending the basic capabilities with custom data structures, specialised processing pipelines, and leveraging additional hardware resources such as lookup tables (LUTs) and parallel processing~\cite{LUT}. Given that P4 offers minimal support for training machine learning models, we can only deploy pre-trained models. For the purpose of network intrusion detection, we used a pre-trained DT model, as proposed and evaluated in \cite{lightbulb}, to illustrate the feasibility of implementing an efficient in-band NIDS that can work in high throughput environments. This pre-trained Decision Tree classifier was selected because of its efficient design, which delivers strong classification performance across several publicly available benchmark datasets, making it a practical choice within P4’s constraints for high-performance network traffic analysis.


\begin{listing}[!t]
    \centering
\begin{lstlisting}[caption=Decision Tree Implementation in P4, language=c, label={lst:DT}]
...
bit<32> IN_PKTS;//1
bit<16> MIN_IP_PKT_LEN;//5
...
bit<16> L4_DST_PORT;//17
bit<16> L4_SRC_PORT;//18
received_packet_counter.read(IN_PKTS,current_flow_id);
...
dstport_register.read(L4_DST_PORT,current_flow_id);
srcport_register.read(L4_SRC_PORT,current_flow_id);
if(TCP_WIN_MAX_OUT <= 26865){
    if(NUM_PKTS_1024_TO_1514_BYTES <= 120){
        if(IN_PKTS <= 45){
            if(MIN_TTL <= 36){
                if(TCP_WIN_MAX_OUT <= 2){
                malicious_flag_register.write(current_flow_id,1);
                } else {
                   ...
                } else {
                malicious_flag_register.write(current_flow_id,0);
...                 
\end{lstlisting}
\end{listing}

The adopted DT classifier is a NetFlow-based NIDS that uses NetFlow fields to form the conditions within \textit{if-else} statements. Given P4's limited expressiveness and control flow capabilities, it is notable that the language still supports the use of multiple \textit{if-else} statements, allowing for the implementation of this DT model.
The main part of the DT classifier code, as implemented in our P4 program, is shown in code listing~\ref{lst:DT}. In this implementation, NetFlow fields, read from P4 registers, are distributed across the \textit{if-else} statements based on the conditions defined by the pre-trained DT model, as proposed in \cite{lightbulb}. Despite P4's constraints, this approach effectively leverages the language's capabilities to achieve the desired network intrusion detection.

\section{Evaluation and Results} \label{Evaluation and Results}

In this study, we conducted two sets of experiments. The first set involved using an emulated network environment to validate the proof of concept and compare the performance of our model with previous solutions that were evaluated solely in emulated environments. In the second phase, we used a real-world P4-compatible hardware setup to assess the system’s performance under practical conditions, offering a more accurate measure of its effectiveness and scalability evaluated on a real-world P4-compatible hardware setup.

\subsection{Emulated Environment}
\subsubsection{Experimental Setup}
\begin{figure}[!t]
	\centering
	\includegraphics[width=0.8\linewidth]{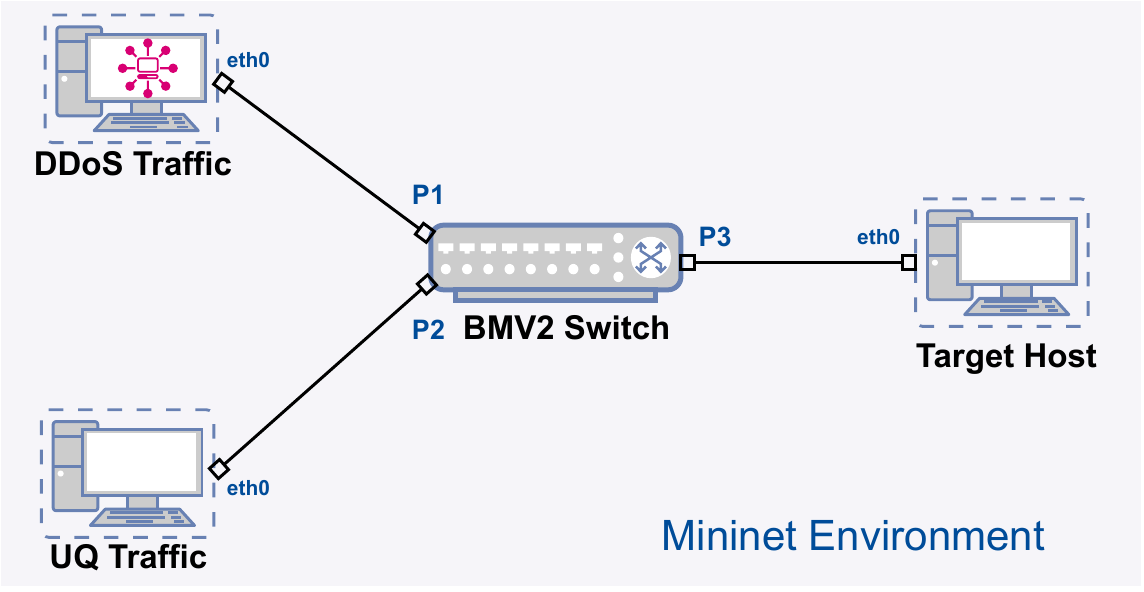} 
	\caption{Emulated network setup}
	\label{fig:Emulated network diagram}
\end{figure}

Figure~\ref{fig:Emulated network diagram} illustrates the emulated network used in this study. For our emulated network setup, we utilised a Dell Optiplex 7070 machine equipped with an Intel Core i7-9700 CPU operating at 3.00 GHz and 24GB of DDR4 2666 MHz RAM. The host operating system was a Desktop Ubuntu 22.04 (kernel 6.5.0-25-generic).
All experiments were conducted on a VirtualBox VM configured with 6 virtual CPU core (100\% capacity) and 8 GB RAM, running Desktop Ubuntu 16.04 (Kernel 4.4.0-150-generic). The network elements, including clients and servers for both benign and attack traffic, as well as the target host, were created using the Mininet network emulator~\cite{mininet}. Mininet is the standard emulation tool employed by the P4 consortium to test P4 program functionalities.

We implemented our DT classifier in the P4 software switch, BMV2~\cite{BMV2}, based on P4-16~\cite{P4-16} with P4-Utils~\cite{p4utils}. 
For the performance and overhead evaluation of the implemented modules, the NetFlow exporter and DT-based classifier we used iperf~\cite{iperf3}.
%



\subsubsection{Performance Evaluation}
To evaluate the scalability of the proposed solution, we started by setting a baseline, measuring the system’s initial performance metrics. Following this, we applied our proposed algorithms to observe any changes in performance.
Accordingly, Our experiments included three different scenarios to thoroughly evaluate the system's behaviour:
\begin{enumerate}
\item \textbf{Basic Port Forwarding:} In this scenario, only a simple port forwarding setup is implemented. All packets arriving at either port P1 or P2 are forwarded to port P3, and vice versa, with no additional processing.

\item  \textbf{NetFlow Generation Enabled:} In this case, we enabled NetFlow generation for the packets being forwarded between switch ports.

\item \textbf{DT-based NIDS Enabled:} Finally, in this scenario, we activated the DT-based NIDS, which is responsible for classifying network traffic flows as either benign or potentially malicious.
\end{enumerate}
Throughout all three scenarios, the CPU utilisation consistently remained at 0.3\%, indicating that the added NetFlow exporter and DT-based classifier modules impose a minimal load on the processor when emulated on the BMv2 switch. This suggests that our additional modules scale efficiently without overloading the system’s CPU. 

For comparison, we used two recent implementation of similar approaches in P4.
As highlighted in the related work, FlowStalker~\cite{castanheira2019flowstalker}  represents the current state-of-the-art in flow extraction schemes within P4, and we use it to benchmark the performance of our NetFlow extraction module.  

\begin{figure}[!t]
    \centering
    \includegraphics[width=0.8\columnwidth]{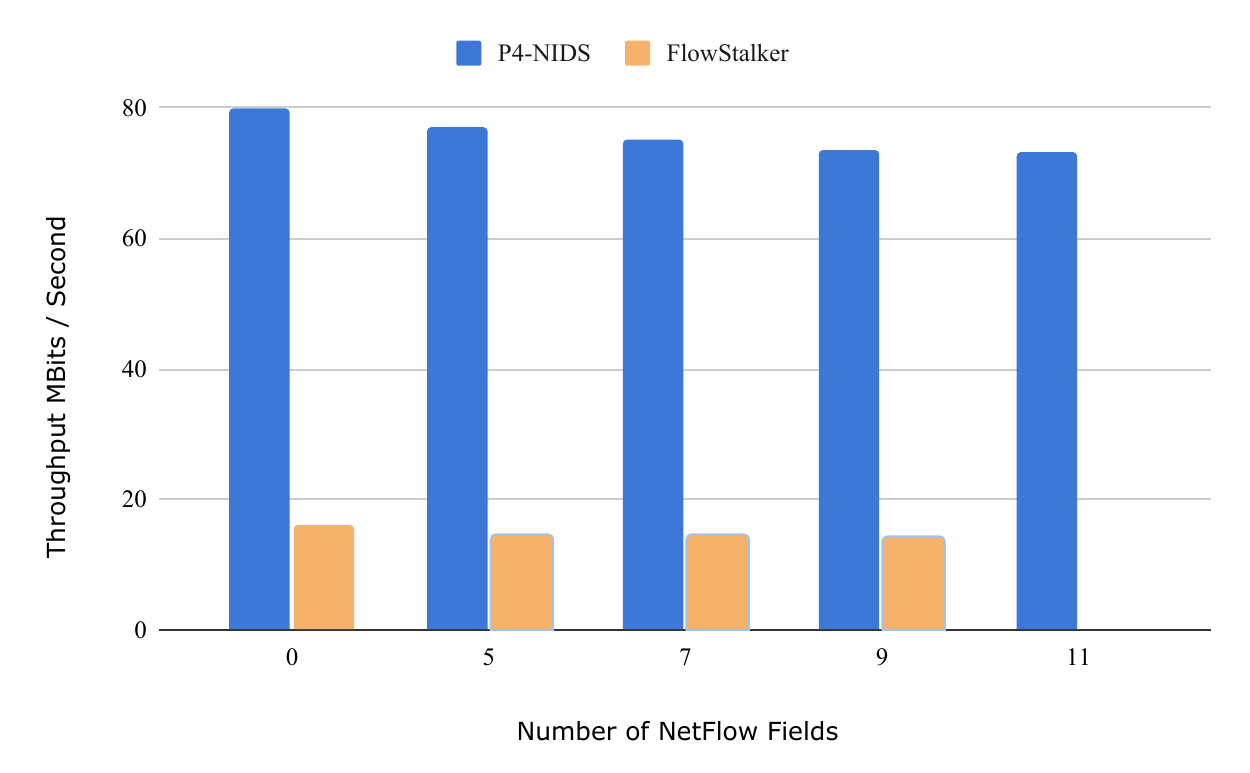} 
    \caption{Throughput comparison of P4-NIDS and FlowStalker by NetFlow features}
    \label{fig:Throughput}
\end{figure}

Figure~\ref{fig:Throughput} shows the throughput of our proposed NetFlow exporter for different number of features, side by side with that of FlowStalker~\cite{castanheira2019flowstalker}.
For this comparison we used similar computational resources and both solutions are evaluated in BMv2 switch.
As can be seen, under the similar condition, the throughput of our solution is about 4 times more than that of FlowStalker~\cite{castanheira2019flowstalker}.

\begin{figure}[!b]
    \centering
    \includegraphics[width=0.8\columnwidth]{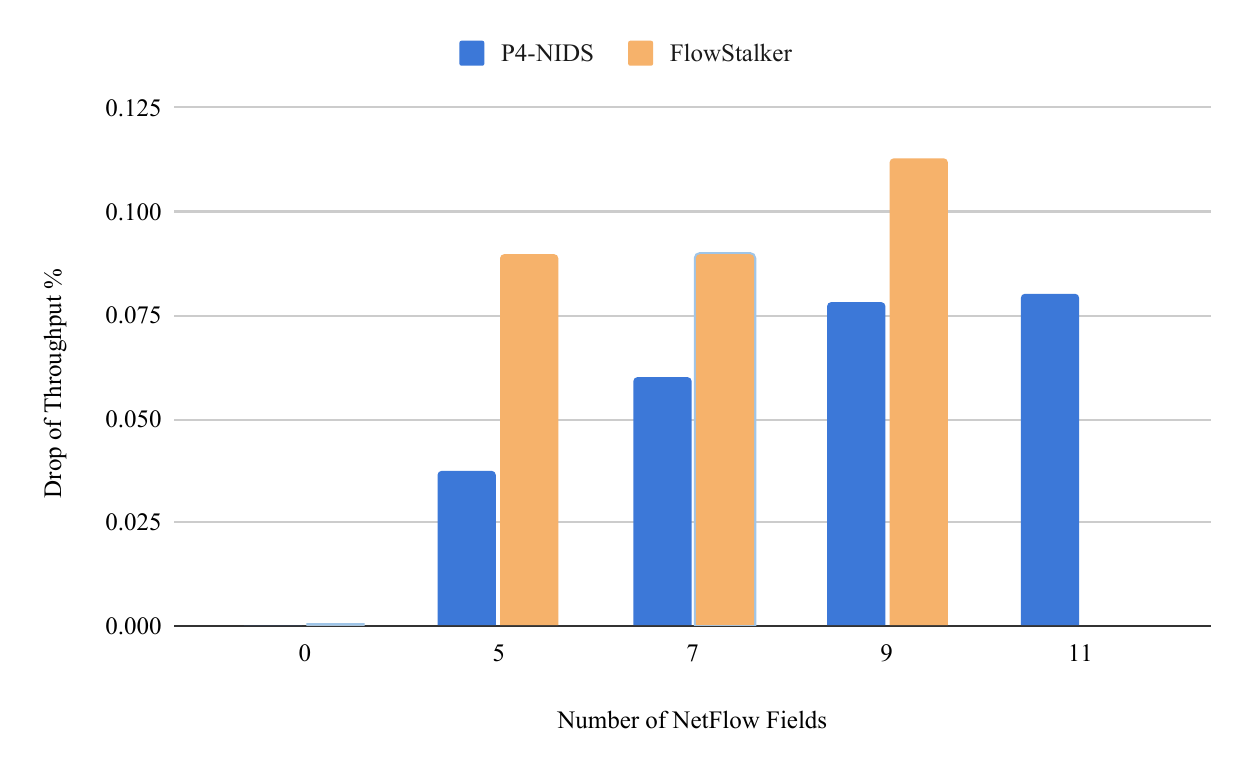} 
    \caption{Throughput drop of P4-NIDS and FlowStalker by NetFlow features}
    \label{fig:DropOfThroughput}
\end{figure}

Similarly, Figure~\ref{fig:DropOfThroughput} shows the drop in throughput due to increasing the number of exported NetFlow fields for both our solution and FlowStalker.
As can be seen, for the equal number of NetFlow fields, our solution shows between 25\% to 50\% less throughput drop than those of FlowStalker.


We also evaluated our solution's classification performance using two widely recognised NIDS benchmarking datasets: CIC-IDS-2018~\cite{CIC-2018} and UNSW-NB155~\cite{UNSW-NB15}. To assess the effectiveness of our DT-based classifier, we compared its F1-score with that of Leo, a recent solution implementing a similar pre-trained DT classifier in P4 for network intrusion detection.
Table~\ref{tab:leo-compare} presents the F1-scores of our solution alongside Leo’s. Our DT-based classifier demonstrates significantly better performance on the UNSW-NB15 dataset. For the CIC-IDS dataset, we used CIC-IDS-2018, an improved version of CIC-IDS-2017 that addresses prior issues. While this comparison isn’t fully equivalent, our results show that our classifier’s F1-score outperforms Leo’s SRAM-based version and is only slightly lower than Leo’s TCAM-based implementation.

\begin{table}[!t]
  \centering
  \caption{Classification performance (F1-score) on different datasets}
    \begin{tabular}{|l|c|c|}
    \hline
    \multicolumn{1}{|c|}{\multirow{2}[4]{*}{Solution}} & UNSW-NB15 & CIC-2018/2017 \\
\cline{2-3}          & F1-score & F1-score \\
    \hline
    Leo-SRAM & 92.00 & 93.00 \\
    \hline
    Leo-TCAM & 93.00 & 99.00 \\
    \hline
    Our solution (BMv2) & 99.76 & 98.09 \\
    \hline
    \end{tabular}%
  \label{tab:leo-compare}%
\end{table}%

%
%
%


%

%

\begin{figure}[!b]
	\centering
	\includegraphics[width=1\columnwidth]{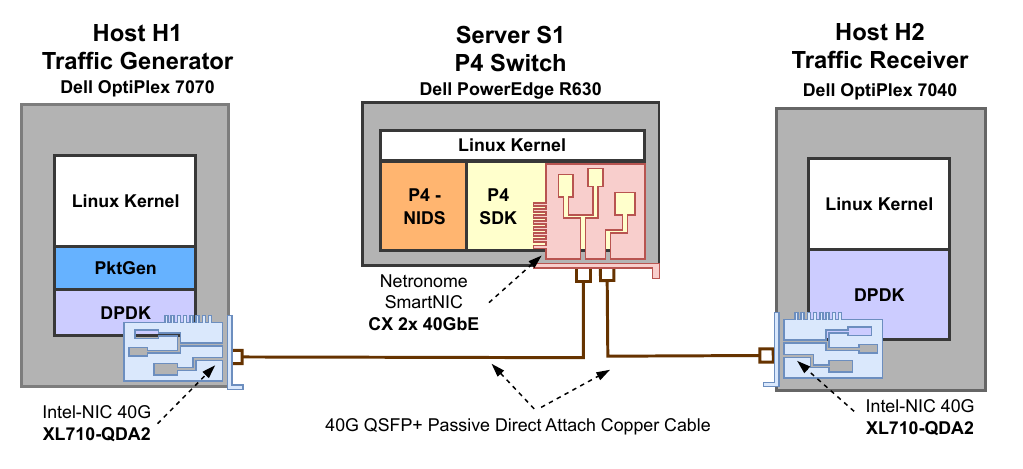} 
	\caption{Hardware experiment setup}
	\label{fig:Hardware-setup}
\end{figure}

\subsection{P4-compatible Hardware}
\subsubsection{Experiment Setup}
Our hardware-based testbed consists of Host H1, Server S1 and Host H2, as shown in Figure~\ref{fig:Hardware-setup}.

Host H1: a Dell OptiPlex 7070 machine running an Intel Core i7-9700 CPU with a clock rate of 3.00 GHz, and 24GB RAM (DDR4 2666 MHz). The host operating system is a Desktop Ubuntu 22.04 (kernel 6.5.0-25-generic). At Host H1, we enable the Pktgen-DPDK~\cite{pktgenDPDK} to achieve high-performance traffic replay and generation.

Server S1: a Dell PowerEdge R630 machine running an Intel Core E5-2630 CPU with a clock rate of 2.40 GHz. The host operating system is a Desktop Ubuntu 18.04 (kernel 4.15.0-142-generic). A Netronome CX Agilio 2x40GbE SmartNIC with a deployed firmware containing P4-NIDS program is inserted in the Server. This host is used to power, flash firmware, and configure the card.

Host H2:  a Dell OptiPlex 7040 machine running an Intel Core i7-6700 CPU with a clock rate of 3.40 GHz, and 15GB RAM.  The host operating system is a Desktop Ubuntu 18.04 (kernel 5.4.0-84-generic). At Host H2, DPDK~\cite{dpdk} is used to receive the traffic and measure the throughput.

%

\subsubsection{Performance Evaluation}
In order to evaluate the resource utilisation of P4-NIDS, we use the same three scenarios explored in the emulated environment for the performance evaluation namely Basic Port Forwarding, NetFlow Generation Enabled and DT-based NIDS Enabled. For this purpose, we measured memory utilisation and throughput drop for each scenario.

For memory utilisation measurement, we ran these scenarios in Program Studio~\cite{netronome_p4c_sdk}, which is an integrated development environment (IDE) designed for developing and debugging networking applications on Netronome's Network Flow Processors (NFPs)~\cite{netronome_nfp_toolchain}. It offers a graphical interface that aligns with the standard look and feel of Microsoft Windows, allowing developers to customise their workspace to fit personal workflows and preferences. It generated the memory usage report providing comprehensive details about memory usage across various types of memories within the system, as explained below.
\begin{itemize}
    \item \textbf{CTM (Cluster Target Memory):} \\Fast memory is typically used for storing packet data or scratch space during packet processing.
    \item \textbf{IMEM (Instruction Memory):} \\Memory used to store the program's instructions. It indicates how much code is loaded in each memory island.
    \item \textbf{EMEM (External Memory): } \\Memory outside the core processing area, generally used for larger buffers or data not requiring fast access. 
    \item \textbf{CLS (CLS Memory):} \\Memory used by the local packet processor, generally for storing state information during packet processing.
    \item \textbf{LM (Local Memory):} \\Refers to memory used by the micro-engines (ME) in each island.
    \item \textbf{EMEM\_CACHE:} \\Cache memory for EMEM, used to speed up access to external memory.
\end{itemize}
Table~\ref{tab:memory-usage} presents the comparison of memory utilisation between the above mentioned three scenarios.
As can be seen, CTM and IMEM remain the same across all scenarios, while in the case of EMEM, CLS, LM and EMEM CACHE  Memory slightly increases in P4-NetFlow and P4-NIDS compared to Basic Forwarding. 
Given the minimal memory increase, which remains within the kilobyte range, these results clearly indicate the lightweight nature of P4-NIDS and its suitability for large-scale network applications.

\begin{table}[!t]
\footnotesize
  \centering
  \caption{Resource utilisation on Netronome SmartNIC (CX 2x40GbE)}
  \label{tab:memory-usage}%
    \begin{tabular}{|c|c|c|c|}
    \hline
\multicolumn{1}{|c|}{\textbf{Memory Parts}} & \textbf{Basic Forwarding} & \textbf{P4-NetFlow} & \textbf{P4-NIDS} \\
\multicolumn{1}{|c|}{} & \textbf{(MB)} & \textbf{(MB)} & \textbf{(MB)} \\
    \hline
    \multicolumn{1}{|l|}{\textbf{Cluster Target Memory}} & 0.828580  & 0.828580 &  0.828580\\
    \hline
    \multicolumn{1}{|l|}{\textbf{Instruction Memory}} & 0.783440 & 0.783440  & 0.783440 \\
    \hline
    \multicolumn{1}{|l|}{\textbf{External Memory}} & 965.367812 & 969.89096  & 969.89256 \\
    \hline
    \multicolumn{1}{|l|}{\textbf{Classifier Memory}} & 0.098848 & 0.147232  & 0.151552 \\
    \hline
    \multicolumn{1}{|l|}{\textbf{Local Memory}} & 0.128844 & 0.184140  & 0.184140 \\
    \hline
    \multicolumn{1}{|l|}{\textbf{Cache Memory for EMEM}} & 0.561528 & 1.118592  & 1.118592 \\
    \hline
    \end{tabular}%
\end{table}


\begin{table}[!b]
\centering
\caption{PKT-GEN Bandwidth and Throughput Data}
\begin{tabularx}{\linewidth}{|c|c|X|X|X|X|}
\hline
\footnotesize \textbf{Bandwidth \%} & \small \textbf{Offered Load} & \multicolumn{2}{c|}{\small \textbf{Pkt Size=1024Bytes}} & \multicolumn{2}{c|}{\small \textbf{Pkt Size=64Bytes}} \\ 
\cline{3-6}
 & \small \textbf{Gbits/s} & \makecell{\scriptsize \textbf{Sent Pkt} \\ \small \textbf{(MPPS)}} & \makecell{\scriptsize \textbf{Throughput} \\ \small \textbf{(Gbits/s)}} & \makecell{\scriptsize \textbf{Sent Pkt} \\ \small \textbf{(MPPS)}} & \makecell{\scriptsize \textbf{Throughput} \\ \small \textbf{(Gbits/s)}} \\ 
\hline
10\% & 4 & 0.5 & 4 & 5.929 & 4 \\ 
\hline
20\% & 8 & 1 & 8 & 7.754 & 5.267 \\ 
\hline
30\% & 12 & 1.5 & 12 & 7.755 & 5.288 \\ 
\hline
40\% & 16 & 2 & 16 & 7.701 & 5.289 \\ 
\hline
50\% & 20 & 2.5 & 20 & 7.699 & 5.284 \\ 
\hline
60\% & 24 & 3 & 24 & 7.793 & 5.282 \\ 
\hline
70\% & 28 & 3.188 & 25.5 & 7.771 & 5.286 \\ 
\hline
80\% & 32 & 3.228 & 25.824 & 7.834 & 5.258 \\ 
\hline
90\% & 36 & 3.243 & 25.94 & 7.791 & 5.291 \\ 
\hline
100\% & 40 & 3.227 & 25.819 & 7.717 & 5.302 \\ 
\hline
\end{tabularx}
\label{tab:PKE-GEN-MPPS}%
\end{table}

%
\begin{figure*}[!t]
	\centering
	\includegraphics[width=1\linewidth]{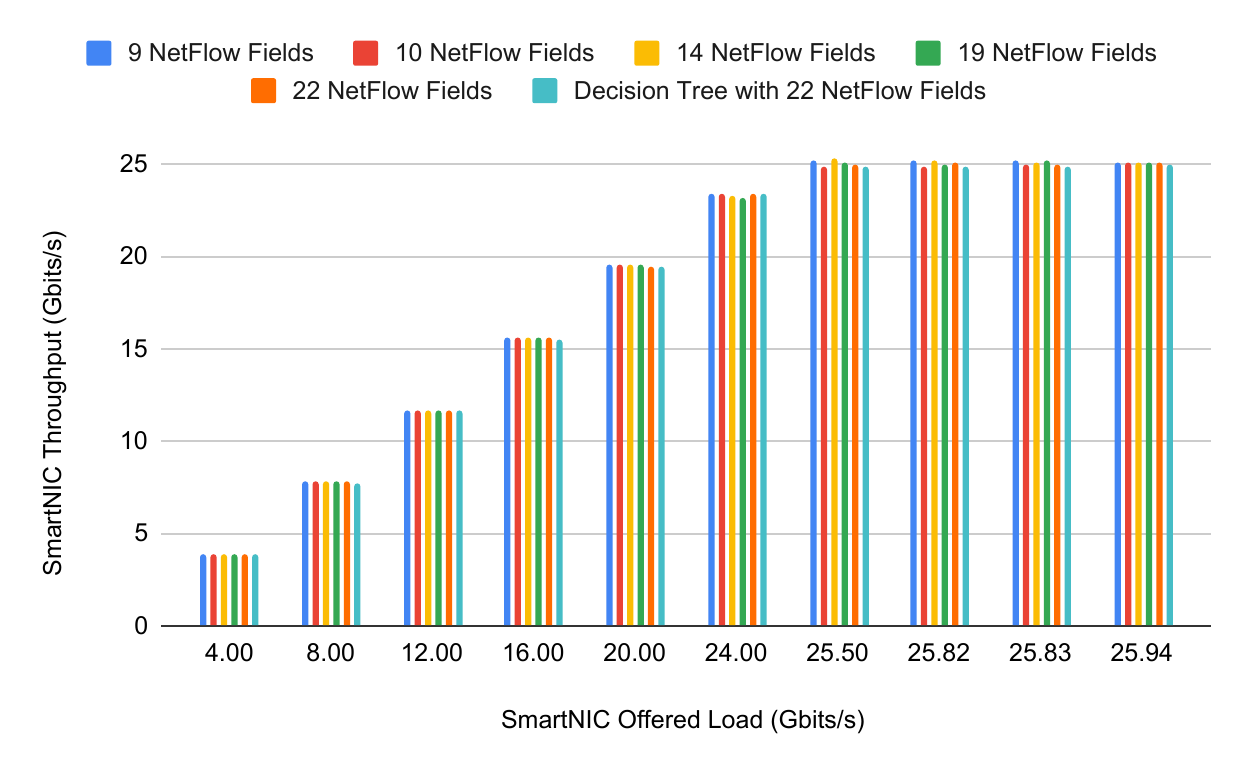} 
	\caption{Thoughput Vs. OfferedLoad}
	\label{fig:Thoughput-OfferedLoad-Hardware}
\end{figure*}

For the throughput degradation analysis, we used Pktgen-DPDK~\cite{pktgenDPDK}, powered by DPDK, to benchmark the performance metrics. Different bandwidth limits were applied in Pktgen-DPDK on Host1 to evaluate throughput under various load conditions. Since our hardware supports a theoretical maximum bandwidth of 40 Gbps, we expected an output of e.g. 4 Gbps at 10\% bandwidth and 40 Gbps at full capacity. 
However, as shown in Table~\ref{tab:PKE-GEN-MPPS}, the Intel NIC hardware limitations restrict the maximum achievable traffic rate from Pktgen to about 25.940 Gbps at 100\% bandwidth. 
The throughput for smaller packet sizes is significantly lower due to the higher number of packets processed by the hardware. For example, while the throughput for a 40 Gbps offered load is 25.819 Gbps with a 1024-byte packet size, it drops to only 5.302 Gbps with a 64-byte packet size.

Considering these constraints, we evaluated the throughput of SmartNIC across five different NetFlow generation scenarios, each with varying feature sets and DT-based NIDS using 22 NetFlow features, under different offered loads.
The experimental results are presented in   Figure~\ref{fig:Thoughput-OfferedLoad-Hardware}.
The findings show that, in all scenarios, SmartNIC throughput closely matches the offered load, demonstrating high performance and minimal impact, particularly in the case of P4-NIDS (i.e., the DT-based classifier). Additionally, for a fixed SmartNIC offered load, throughput remains nearly constant, regardless of the number of NetFlow fields or the inclusion of DT-based NIDS.


\section{Conclusion}\label{Conclusion}
This study addresses the challenge of achieving high-performance and efficient data-plane implementation for network monitoring and machine learning-based intrusion detection. Unlike prior solutions that explore alternative approaches, our proposed method delivers comparable or superior performance to the state-of-the-art, while maintaining minimal computational overhead.

Given the computational constraints of SDN data planes and the programming limitations of P4—such as restricted memory access, limited control flow, and arithmetic operations~\cite{hang2019programming}—not all machine learning models can be implemented in P4. To address this, we adopted an efficient decision tree-based traffic classifier previously proposed in the literature and implemented it in P4. 

The results of the evaluation of our solution on a P4-compatible hardware setup indicate that P4-NIDS achieves high efficiency and performance. It exhibits minimal memory overhead, remains scalable for large networks, and delivers stable throughput under varying loads, confirming its viability for real-world deployment in high-speed networking environments.
Our proposed solution for NetFlow generation and export demonstrates throughput at least four times higher than the best alternative solution. Moreover, it achieves 25\% to 50\% less throughput degradation as the number of exported NetFlow fields increases, outperforming the current state-of-the-art.

These findings highlight the lightweight nature of our approach, ensuring it is well-suited for large-scale network applications. Additionally, the decision tree-based intrusion detection system (P4-NIDS) integrated within the solution effectively balances computational efficiency with high classification accuracy, even under demanding traffic conditions.

In conclusion, the proposed solution sets a new benchmark for implementing network monitoring and intrusion detection within the P4 data plane, addressing key challenges of computational constraints and scalability. Its superior performance, low overhead, and compatibility with existing hardware make it an ideal choice for modern high-performance networking environments, paving the way for more efficient and robust SDN-based security solutions.


\bibliographystyle{unsrt}
\bibliography{main}


\end{document}